\documentclass[%
reprint,onecolumn,11pt,
amsmath,amssymb,
prb,
]{revtex4-1}

\usepackage{graphicx}
\usepackage{dcolumn}
\usepackage{bm}

\begin{document}

\title{High resolution nuclear magnetic resonance spectroscopy of highly-strained quantum dot nanostructures}

\author{E. A. Chekhovich$^1$, K. V. Kavokin$^2$, J. Puebla$^1$, A. B. Krysa$^3$, M. Hopkinson$^3$, A. D. Andreev$^4$, A. M. Sanchez$^5$, R. Beanland$^5$,  M. S. Skolnick$^1$, A. I. Tartakovskii$^1$}

\affiliation{$^1$Department of Physics and Astronomy, University of Sheffield, Sheffield S3 7RH, UK\\
$^2$A. F. Ioffe Physico-Technical Institute, 194021, St. Petersburg, Russia\\
$^3$Department of Electronic and Electrical Engineering, University of Sheffield, Sheffield S1 3JD, UK\\
$^4$Hitachi Cambridge Laboratory, Cavendish Laboratory, Madingley Road, Cambridge CB3 0HE, UK\\
$^5$Department of Physics, University of Warwick, Coventry CV4
7AL, UK}

\date{\today}


\begin{abstract}
\end{abstract}

\maketitle


\textbf{Much new solid state technology for single-photon sources
\cite{Salter}, detectors \cite{vanKouwen,Reimer}, photovoltaics
\cite{Tian} and quantum computation \cite{Brunner,Xu1} relies on
the fabrication of strained semiconductor nanostructures.
Successful development of these devices depends strongly on
techniques allowing structural analysis on the nanometer scale.
However, commonly used microscopy
methods\cite{Siverns,Tanaka,Bruls,Wu} are destructive, leading to
the loss of the important link between the obtained structural
information and the electronic and optical properties of the
device. Alternative non-invasive techniques such as optically
detected nuclear magnetic resonance (ODNMR) so far proved
difficult in semiconductor nano-structures due to significant
strain-induced quadrupole broadening of the NMR spectra. Here, we
develop new high sensitivity techniques that move ODNMR to a new
regime, allowing high resolution spectroscopy of as few as 10$^5$
quadrupole nuclear spins. By applying these techniques to
individual strained self-assembled quantum dots, we measure strain
distribution and chemical composition in the volume occupied by
the confined electron. Furthermore, strain-induced spectral
broadening is found to lead to suppression of nuclear spin
magnetization fluctuations thus extending spin coherence times.
The new ODNMR methods have potential to be applied for
non-invasive investigations of a wide range of materials beyond
single nano-structures, as well as address the task of
understanding and control of nuclear spins on the nanoscale, one
of the central problems in quantum information processing
\cite{Khaetskii,Bluhm,Foletti}.}

Most nuclei used in optically active III-V semiconductor
nanostructures possess non-zero quadrupole moments sensitive to
electric field gradients caused e.g. by strain. Due to strong
spectral broadening NMR on quadrupole nuclei is challenging even
for macroscopic samples
\cite{ODell200897,Tang2008227,CMR:CMR20037}. As a result control
of nuclear spins using NMR has only been achieved in strain-free
GaAs/AlGaAs semiconductor quantum dots (QDs)
\cite{GammonScience,MakhoninPRB,MakhoninNatMat}. By contrast
application of similar techniques to widely researched strained
self-assembled quantum dots were limited to large ensembles of
QDs, where quadrupole broadening lead to uncertainty in
interpretation of the effects observed under radio-frequency (rf)
excitation \cite{Flisinski,Cherbunin}.

The new ODNMR spectroscopy technique we present is based on
continuous-wave broadband rf excitation with specially designed
spectral patterns, which for nuclei with spin $I$ in strained
structures provides sensitivity enhancement by a factor of
$(I+1/2)^3$ compared to conventional saturation NMR techniques
previously used in unstrained nano-structures
\cite{GammonScience,MakhoninPRB,MakhoninNatMat}. Such enhancement
(particularly large for high spin nuclei such as $I$=9/2 indium)
brings NMR to a qualitatively new level: it allows high resolution
spectroscopy in individual few-nanometer-sized strained quantum
dots having only $10^4$-$10^5$ nuclear spins which are subject to
large inhomogeneous quadrupole broadening up to 20~MHz. To
demonstrate the capabilities of this technique we use NMR spectra
measured for InP/GaInP and InGaAs/GaAs QDs to estimate several key
structural properties such as material composition, elastic strain
magnitude and distribution in the volume sampled by the electron
wave-function. Furthermore, using related techniques we find that
spectral broadening caused by strain results in enhancement of the
coherence of nuclear spins by a factor of 10 compared with
unstrained structures, leading to coherence times of $\approx 2.5$
ms.

\begin{figure}
\includegraphics{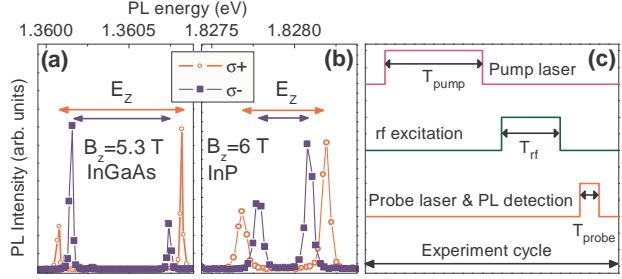}
\caption{\label{fig:SpecScheme} (a) and (b) show typical
photoluminescence spectra for InGaAs/GaAs (a) and InP/GaInP (b)
quantum dots measured in magnetic field $B_z=5.3$ and 6~T,
respectively. Squares (circles) show PL spectra measured for
$\sigma^{-(+)}$ excitation exhibiting differing Zeeman splittings
$E_z$ due to the nuclear spin polarization (anti-)parallel to
$B_z$. (c) shows the timing diagram of the experimental cycle
including optical pump and probe, and rf excitation. We use
$T_{pump}=4\div7$~s, $T_{rf}=4\div6$~s and
$T_{probe}=3\div16$~ms.}
\end{figure}

In what follows we present results for two different types of
strained semiconductor nano-structures: InP/GaInP and InGaAs/GaAs
self-assembled quantum dots [for sample details see Supplementary
Information (SI) Section S1
]. All measurements were performed at $T=4.2$~K, in external
magnetic field $B_z$ normal to the sample surface. Under
excitation with circularly polarized light, nuclear spins become
strongly polarized due to spin transfer from electrons via the
hyperfine interaction \cite{GammonPRL}. The resulting nuclear spin
polarization on the dot is detected in photoluminescence (PL) of
excitons in single QDs as shown in Fig. \ref{fig:SpecScheme}(a)
for InGaAs and in Fig. \ref{fig:SpecScheme}(b) for InP dots in
high magnetic field $B_z>5$~T. Each spectrum consists of an
exciton Zeeman doublet with splitting $E_z$. Detection of changes
in $E_z$ allows measurement of the electron Overhauser shift
\cite{GammonPRL,InPRes} proportional to the degree of nuclear
polarization $P_N$. ODNMR measurements were carried out using the
pump-probe method schematically shown in Fig.
\ref{fig:SpecScheme}(c). The dot is first excited with a
circularly polarized laser pulse of duration $T_{pump}$. After
that the rf excitation is applied in the dark for duration
$T_{rf}$. Finally a short ($T_{probe}$) laser pulse is applied to
measure the PL spectrum and probe the effect of the rf field on
nuclear polarization, which allows NMR signal to be obtained as
the absolute magnitude of the Overhauser shift. More details of
experimental methods can be found in SI Sec.
S2
.

\begin{figure}
\includegraphics{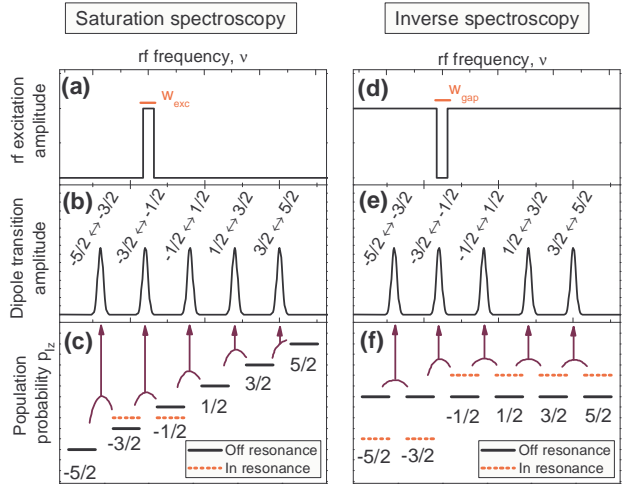}
\caption{\label{fig:InvNMRScheme} Schematics explaining
''saturation'' (a)-(c) and ''inverse'' (d)-(f) NMR spectroscopy
applied to quadrupolar nuclei. The case of an ensemble of spin-5/2
nuclei subject to the same electric field gradient is considered
for clarity. (a) and (d) show rf excitation spectra in the two
spectroscopy methods. Spectra of dipole transitions between the
nuclear spin levels are shown in (b) and (e). (c) and (f) show
population probabilities of the nuclear spin levels. Arrows show
the transitions in the nuclear spectra in (b) and (e)
corresponding to pairs of the spin states coupled by the rf field.
Solid lines show the population probabilities for the case when
the rf maximum in (a) and the gap in (d) are off resonance with
all transitions. Dashed lines show the case when the maximum in
(a) and the gap in (d) are in resonance with $-3/2 \leftrightarrow
-1/2$ transition.}
\end{figure}

The explanation of our method and its comparison with the
''saturation'' NMR techniques applied in strain-free materials
\cite{MakhoninPRB,GammonScience} is given in Fig.
\ref{fig:InvNMRScheme}, using an example of spin $I=5/2$ nuclei.
In an external magnetic field $B_z$ along the $Oz$ axis nuclear
spin levels experience Zeeman shifts $\propto B_z I_z$ determined
by their spin projections $I_z$. The oscillating magnetic field
$B_{rf}$ perpendicular to $Oz$ couples only spin levels with $I_z$
differing by $\pm1$. If the nucleus is subject to an electric
field gradient (EFG) along the $Oz$ axis (e.g. induced by elastic
strain), in addition to the splitting induced by $B_z$, the
energies of spin levels will change \cite{AbrahamBook,QEncycl} by
a value proportional to $I_z^2$ (see SI Sec. S3
), and all dipole active transitions will
have different frequencies, as depicted in Fig.
\ref{fig:InvNMRScheme}(b).

Let us consider an ensemble of nuclei with spins $I=5/2$ all
subject to the same EFG and polarized, so that $I_z=5/2$ levels
have higher population than $I_z=-5/2$. The probabilities
$p_{I_z}$ to find a nucleus with spin $I_z$ will depend on $I_z$
as sketched by the solid lines in Fig. \ref{fig:InvNMRScheme}(c).
The total nuclear spin polarization degree (which is detected
optically) is $P_N=\sum_{k=-I}^{I} p_{k}\times k/I$, so that
$|P_N|\leq 1$. Arrows in Fig. \ref{fig:InvNMRScheme}(c) indicate
the maxima in the NMR spectrum in Fig. \ref{fig:InvNMRScheme}(b)
corresponding to allowed transitions between adjacent pairs of
nuclear spin levels.

In conventional ''saturation'' NMR spectroscopy, radio-frequency
excitation at a frequency $\nu$ or, a distribution of frequencies
with a width $w_{exc}$ is applied [Fig.
\ref{fig:InvNMRScheme}(a)]. $P_N$ will only change in the case
when  $\nu$ is in resonance with a transition between $I_z'$ and
$I_z'+1$ levels, for example, $-3/2\leftrightarrow -1/2$ in Figs.
\ref{fig:InvNMRScheme}(a-c). This occurs as the populations of
these spin levels equalize under sufficiently long resonant rf
excitation [dotted lines in Fig. \ref{fig:InvNMRScheme}(c)], which
at the same time has no effect on populations of the other spin
levels. As a result, the overall change in nuclear polarization is
small, making the resonance difficult to detect.

Very major enhancement of the changes in $P_N$ can be achieved by
using an alternative approach developed in this work [see Fig.
\ref{fig:InvNMRScheme}(d)]: we use broad band excitation with a
continuum spectrum containing a gap of width $w_{gap}$. As the rf
excitation spectrum in Fig. \ref{fig:InvNMRScheme}(d) is an
inversion of that in Fig. \ref{fig:InvNMRScheme}(a), we introduce
the term ''\textit{inverse}'' spectroscopy. The effect of such
excitation on the populations $p_{I_z}$ is demonstrated in Fig.
\ref{fig:InvNMRScheme}(f). If the gap is out of resonance with all
transitions, all $p_{I_z}$ are equalized (solid lines) and nuclear
spin polarization is completely erased ($P_N=0$). If, however, the
gap is in resonance with the $I_z'\leftrightarrow I_z'+1$
transition, i. e. one of the transitions is not excited, the
equalization of populations occurs independently for two groups of
levels with $I_z\le I_z'$ or $I_z\ge I_z'+1$ (dotted lines). Thus
the resonance condition for one of the transitions affects the
populations of all states including $I_z=\pm I$ states, which give
the largest contribution to $P_N$. In the experiment, the
''inverse'' NMR spectrum is obtained by scanning the central
frequency $\nu$ of the gap, while $w_{gap}$ is chosen to control
the balance between the spectral resolution and sensitivity (NMR
signal amplitude). It is possible to show (see SI Sec. S4
) that for ''inverse'' method the enhancement of
the changes in nuclear polarization exceed $(I+1/2)^3$ ($=$125 for
spin $I=9/2$) compared to the saturation NMR method in Fig.
\ref{fig:InvNMRScheme}(a). This is a significant improvement
compared to the existing ''population transfer'' technique where a
maximum enhancement of $2 I$ can be achieved \cite{CMR:CMR20037}.

\begin{figure*}
\includegraphics{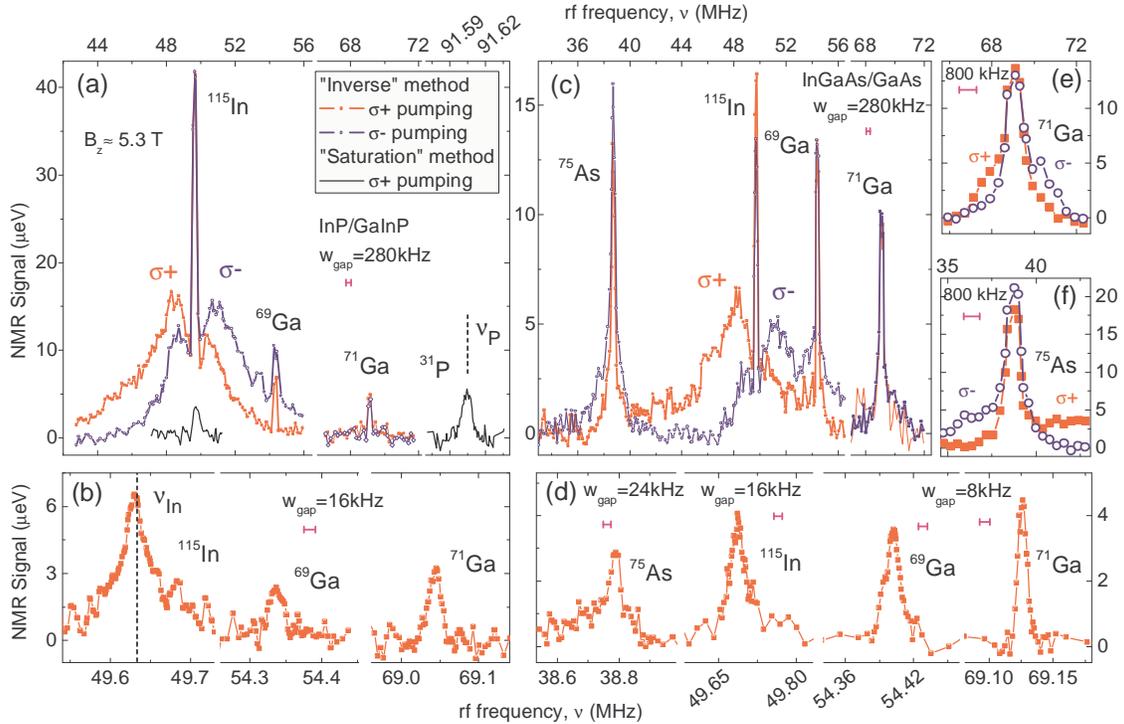}
\caption{\label{fig:NMRSpec} ODNMR spectra measured at
$B_z\approx$5.3~T in self-assembled QDs using the ''inverse''
method with $\sigma^+$ (red curves) or $\sigma^-$ (blue curves)
optical pumping and using the ''saturation'' method with
$\sigma^+$ pumping (black curves). (a) and (b) show ODNMR spectra
for InP/GaInP dots measured with resolution $w_{gap}=$280 kHz in
(a) and 16 kHz in (b), and applying ''saturation'' spectroscopy
with a width $w_{exc}=$450 kHz for $^{115}$In and single-frequency
excitation for spin-1/2 $^{31}$P in (a). (c)-(f) show ODNMR
spectra for InGaAs/GaAs dots measured with resolution
$w_{gap}=$280 kHz in (c), 8-24 kHz in (d), and 800 kHz in (e,f).
Vertical line in (b) shows indium frequency
$\nu_{In}\approx49633$~kHz corresponding to unstrained InP.}
\end{figure*}

Fig. \ref{fig:NMRSpec} shows a set of ODNMR spectra measured on
single InP (a,b) and InGaAs (c-f) QDs at $B_z\approx5.3$~T. In the
nominally InP dots grown in GaInP barriers, the ''inverse''
technique allows to resolve contributions from quadrupole nuclei
$^{115}$In, $^{69}$Ga and $^{71}$Ga within the volume probed by a
single electron. In Fig. \ref{fig:NMRSpec}(a) the $^{115}$In peak
dominating the spectrum, consists of a sharp central line
[corresponding to $-1/2\leftrightarrow 1/2$ central transition
(CT)] with amplitude $\sim$40~$\mu$eV at $\sim$49.7~MHz, and two
broad bands of satellite transitions (STs) to lower and higher
frequencies each stretching up to $\pm$7~MHz. These sidebands are
due to strain-sensitive, quadrupole split transitions between spin
levels with $|I_z|>1/2$. The relative amplitudes of the side-bands
reflect the initial alignment of the nuclear spins by circularly
polarized excitation: the high (low) frequency band has a higher
(lower) intensity for $\sigma^-$ ($\sigma^+$) excitation. By
contrast, with the saturation method applied to $^{115}$In nuclei
a weak line with amplitude $\sim$3~$\mu$eV [see Fig.
\ref{fig:NMRSpec}(a)] can be detected only for rf excitation width
$w_{exc}=450$~kHz or larger, i.e. at low resolution. For spin-1/2
$^{31}$P nuclei unaffected by quadrupole effects the saturation
method with monochromatic rf excitation reveals a single line with
a width of $\sim$8~kHz at $\nu_P\approx$91605 kHz.

The ODNMR spectra of a InGaAs/GaAs QD measured using the
''inverse'' method ($w_{gap}$=280 kHz) are shown in Fig.
\ref{fig:NMRSpec}(c). Here the central transitions have similar
amplitudes for the four isotopes present in the dot, revealing
significant substitution of indium by gallium. Satellite
transitions are observed only for spin-$9/2$ $^{115}$In and have
similar widths to the case of InP dots. In order to observe these
transitions for spin-$3/2$ nuclei we have carried out ''inverse''
NMR measurements with a larger $w_{gap}$=800 kHz as shown in Figs.
\ref{fig:NMRSpec}(e) and (f) for $^{71}$Ga and $^{75}$As
respectively. For $^{71}$Ga the spectral range where STs are
observed is within $\pm$2.5~MHz on both sides of the central line,
whereas it is significantly broader for $^{75}$As due to its
larger quadrupole moment $Q$.

The structure of the central transitions is affected by strain
only via weak second-order quadrupole interaction \cite{QEncycl}.
It is measured using ''inverse'' NMR with a smaller
$w_{gap}$=8$\div$24 kHz to provide higher resolution. The smallest
linewidth of $\sim$8~kHz is observed for $^{71}$Ga in InGaAs QDs
while more sensitive to quadrupole interaction $^{75}$As as well
as $^{115}$In in both materials have linewidths of $\sim$40~kHz.
Broadening of the CT line as well as its shift with respect to
resonance frequency in unstrained bulk material [shown by vertical
line in Fig. \ref{fig:NMRSpec}(b) for $^{115}$In in InP] appear if
electric field gradient direction deviates from that of magnetic
field and thus reflect the distribution of elastic strain
directions within the volume probed by the electron.

Quadrupole effects are insensitive to isotropic (hydrostatic)
strain, but can provide information on biaxial and shear strains.
For a given electron wavefunction and strain distribution within
the QD it is possible to calculate an ''inverse'' NMR spectrum
that can be directly compared with experiment. Since detailed
quantum dot modeling \cite{MlinarDotModeling} is beyond the scope
of this work we limit our discussion to some quantitative
estimates. For example, it is observed in Figs. \ref{fig:NMRSpec}
(a,c) that satellite transitions $I_z'\leftrightarrow I_z'+1$ of
$^{115}$In corresponding to different $I_z'$ are not resolved.
This signifies a strong variation of the quadrupole shifts over
the volume of the dot resulting from the variation of elastic
strain. In particular, it can be seen that ST bands have non-zero
amplitudes at the CT frequency, implying that for some nuclei the
quadrupole splitting is zero. This can be explained by relaxation
of the biaxial strain near the center of the dot as predicted by
model calculations \cite{GrundmannStrain,AndreevStrain}. On the
other hand, the maximum values of the biaxial strain can be
readily estimated from the maximum frequency shifts observed for
the STs. For the spin-3/2 $^{71}$Ga, the largest shift of the STs
from the CT in InGaAs/GaAs QDs is $\sim$2.5~MHz [Fig.
\ref{fig:NMRSpec}(e)]. Using the values of the EFG-elastic tensor
measured in bulk GaAs \cite{SundforsSTensor} this allows the
maximum biaxial strain with principle axis along $Oz$ to be
estimated as $|\epsilon_b|\sim6\%$. This is somewhat smaller than
the maximum $|\epsilon_b|\sim9\div15\%$ predicted for interfacial
regions of InAs dots of different shapes \cite{AndreevStrain},
possibly a signature of ''smoothing'' of the interface between the
dot and the barrier due to interdiffusion. Using NMR on $^{115}$In
we find in a similar way the maximum strain for InP/GaInP dots of
$|\epsilon_b|\sim5\%$. (See details in SI Sec. S6
).

The NMR data in Fig. \ref{fig:NMRSpec} can be used to estimate
relative gallium and indium concentrations $\rho_{Ga}$ and
$\rho_{In}$. For InGaAs dots in Fig. \ref{fig:NMRSpec}(c) similar
CT signals are observed for $^{71}$Ga and $^{115}$In isotopes.
However, we need to take into account, that for the $^{115}$In
isotope the NMR signal is enhanced due to the larger nuclear
magnetic moment of indium
$\mu_{(^{115}In)}/\mu_{(^{71}Ga)}\approx2$, and its larger natural
abundance ($\sim$96\%) compared to 40\% for $^{71}$Ga. In this way
we estimate $\rho_{In}\approx20\%$ and $\rho_{Ga}\approx80\%$ for
InGaAs dots. Similar estimates for InP dots give
$\rho_{In}\approx65\%$ and $\rho_{Ga}\approx35\%$. These estimates
are further confirmed by more detailed measurements and analysis
presented in SI Sec. S5
.

The spectral widths exceeding 10~MHz observed for spectra in Fig.
\ref{fig:NMRSpec} are due to inhomogeneous quadrupole broadening.
However, each nuclear spin transition has a finite intrinsic
linewidth $w_{nuc}$ determined by fluctuations of the local
fields, created by neighboring nuclear spins. In unstrained
structures $w_{nuc}$ can be determined by spin-echo techniques
\cite{MakhoninNatMat}, using $\pi$ and $\pi/2$ rf-pulses for
coherent manipulation of nuclear spins. In strained structures
uniform excitation of nuclear spins, with spectral dispersion of
several MHz, would require very short rf-pulses, which in turn
require rf-powers that are difficult to achieve with available
hardware \cite{Tang2008227}. However, as we now show, nuclear spin
coherence can be probed using an alternative approach based on
low-power continuous-wave rf excitation with a special spectral
pattern.

\begin{figure}
\includegraphics{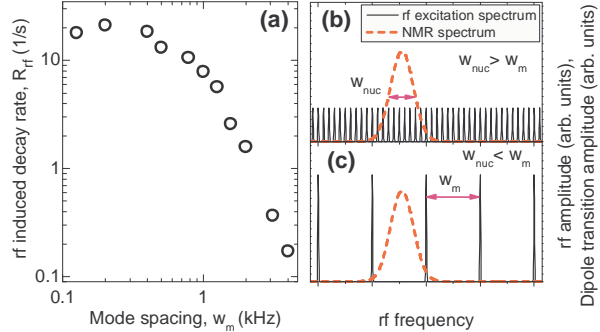}
\caption{\label{fig:ModeSpacing}(a) Experimental dependence of rf
induced nuclear depolarization rate $R_{rf}$ on the spacing $w_m$
between spectral modes of the rf excitation. (b) and (c)
qualitatively explain the effect of the variation of $w_m$ on the
nuclear polarization decay rate (see text). Solid and dashed lines
show schematically spectra of rf excitation and nuclear spin
dipole transition, respectively.}
\end{figure}

This approach is demonstrated in Fig. \ref{fig:ModeSpacing} for
InP QDs. We measure the rate $R_{rf}$ of nuclear spin
depolarization induced by broad-band rf excitation consisting of
equally spaced spectral modes [rf spectra are sketched with solid
lines in Figs. \ref{fig:ModeSpacing} (b,c)]. The measurements are
carried out for different spacings $w_m$ between the modes, while
keeping a constant spectral range 44-47.5~MHz corresponding to the
satellite transitions of In nuclei [see Fig. \ref{fig:NMRSpec}(a)]
and constant total power of the rf excitation.

The strong reduction of $R_{rf}$ with increasing $w_m$ observed in
the experiment [Fig. \ref{fig:ModeSpacing}(a)] is qualitatively
explained in Figs. \ref{fig:ModeSpacing}(b,c). For $w_m<w_{nuc}$
[Fig. \ref{fig:ModeSpacing}(b)] the discrete structure of the
rf-spectrum is averaged out by the broadening of the nuclear spin
transition and each nuclear spin transition is excited by many
modes resulting in fast depolarization. If the mode spacing is
increased ($w_m>w_{nuc}$) there is a high probability that a given
nuclear spin transition is out of resonance with all modes [Fig.
\ref{fig:ModeSpacing}(c)]. Strong suppression of $R_{rf}$ observed
experimentally at $w_m\approx$4~kHz [Fig.
\ref{fig:ModeSpacing}(a)] will take place for very large mode
spacing $w_m\gg w_{nuc}$ when most of the nuclear transitions will
be out of resonance with rf excitation. Model fitting allows an
intrinsic linewidth $w_{nuc}\approx$0.13 kHz to be determined.
Furthermore, analysis of the lineshape of individual nuclear spin
transitions reveals a deviation from a Lorenzian lineshape
indicative of non-exponential dynamics of the transverse nuclear
magnetization (see details in SI Sec. S7
).

From $w_{nuc}$ we can estimate the nuclear spin decoherence time
of indium nuclei as $T_2\approx1/(\pi w_{nuc})\approx2.5$~ms. This
exceeds by nearly an order of magnitude previously reported $T_2$
for quadrupole nuclei in unstrained GaAs/AlGaAs QWs \cite{Sanada}
(0.27 ms) and QDs\cite{MakhoninNatMat} (0.35 ms). This can be
interpreted as a result of large quadrupole shifts, caused by
inhomogeneous elastic strain, that make flips of spin states with
$|I_z|>1/2$ energetically forbidden, partially freezing
fluctuations of the dipole-dipole field. The nuclear spin
decoherence time is likely to be determined by spin flips of the
$-1/2\leftrightarrow 1/2$ transition of both quadrupole nuclei and
phosphorus. Thus, while leading to inhomogeneous spectral
broadening exceeding 10~MHz, quadrupole interaction has the
beneficial effect of narrowing of individual transitions down to
$w_{nuc}\approx$0.13 kHz.

The proposed techniques may become a useful tool in development of
single QDs for experiments on electron and hole spin coherence, as
direct measurements of material properties such as composition and
strain have now been made possible. This will potentially allow a
generation of semiconductor nano-structures with tailored material
properties for these applications. Furthermore, our experiments
reveal very robust coherence properties of nuclear spins in
strained structures, showing that introduction of strain is one of
the ways to achieve a very stable nuclear spin bath, which is of
potential significance for enhancing electron and hole coherence
on the nanoscale. Finally, we note that these techniques may be
applicable in a wide range of NMR detection schemes and structures
beyond quantum dots. One of the challenges will be to develop
nano-NMR techniques (including imaging) sensitive to yet
smaller numbers of nuclei in the solid state environment. \\

{\label{sec:Acknowledgments} \bf ACKNOWLEDGMENTS} This work has
been supported by  the EPSRC Programme Grant EP/G001642/1, ITN
Spin-Optronics and the Royal Society. J.P. has been supported by
a CONACYT-Mexico doctoral scholarship. The authors are thankful to A.~J.~Ramsay and D.~N.~Krizhanovskii for fruitful discussion.\\

{\label{sec:Contributions} \bf AUTHOR CONTRIBUTIONS} A.B.K. and
M.H. developed and grew the samples. A.M.S. and R.B. produced TEM
images of QDs. J.P. processed the samples. E.A.C. and A.I.T.
conceived the experiments. E.A.C. developed new techniques and
carried out the experiments. E.A.C., K.V.K., A.D.A. and A.I.T.
analyzed the data.
E.A.C., A.I.T. and M.S.S. wrote the manuscript with input from all authors.\\

{\label{sec:Information} \bf ADDITIONAL INFORMATION}
Correspondence and requests for materials should be addressed to
E.A.C. (e.chekhovich@sheffield.ac.uk) and A.I.T.
(a.tartakovskii@sheffield.ac.uk)

\renewcommand{\thesection}{S\arabic{section}}
\setcounter{section}{0}
\renewcommand{\thefigure}{S\arabic{figure}}
\setcounter{figure}{0}
\renewcommand{\theequation}{S\arabic{equation}}
\setcounter{equation}{0}

\renewcommand{\citenumfont}[1]{S#1}
\makeatletter
\renewcommand{\@biblabel}[1]{S#1.}
\makeatother

\pagebreak \pagenumbering{arabic}

\section*{SUPPLEMENTARY INFORMATION}

In this document we refer to the figures 1-4 of the main text as
well as to supplementary figures S1-S5.

\quad\\

The document consists of
the following sections:\\
\ref{SI:Samples}. Details of sample structure and growth,\\
\ref{SI:Experiment}. Details of experimental techniques,\\
\ref{SI:NucSpecStrain}. Nuclear spin spectrum in the presence of
strain,\\
\ref{SI:InvNMR}. Calculation of the NMR signal for ''inverse''
spectroscopy,\\
\ref{SI:ChemComp}. Estimation of the chemical composition of the
dots,\\
\ref{SI:StrainEstim}. Estimation of strain in a QD,\\
\ref{SI:ModeSpacing}. Spin coherence in the nuclear spin ensemble
subject to inhomogeneous strain.

\section{\label{SI:Samples}Details of sample structure and growth}

We use InP/GaInP self-assembled quantum dots (QDs) grown by
metal-organic vapor-phase epitaxy (MOVPE) and InGaAs/GaAs quantum
dots grown by molecular beam epitaxy (MBE).  Both samples are not
intentionally doped and have no electric gates.

The InP/GaInP sample \cite{InPDyn,InPRes} was grown in a
horizontal flow quartz reactor using low-pressure MOVPE on (100)
GaAs substrates misoriented by $3^{\circ}$ towards
$\langle111\rangle$. The growth temperature of the GaAs buffer and
bottom Ga$_{0.5}$In$_{0.5}$P layer was 700$^{\circ}$~C. Before
proceeding to the deposition of InP and the Ga$_{0.5}$In$_{0.5}$P
capping layer, the wafer was cooled to 650$^{\circ}$~C. The grown
GaInP layers were nominally lattice matched to GaAs. A low InP
growth rate of 1.1\AA/s and deposition time of 10 seconds were
chosen.

\begin{figure}[b]
\includegraphics{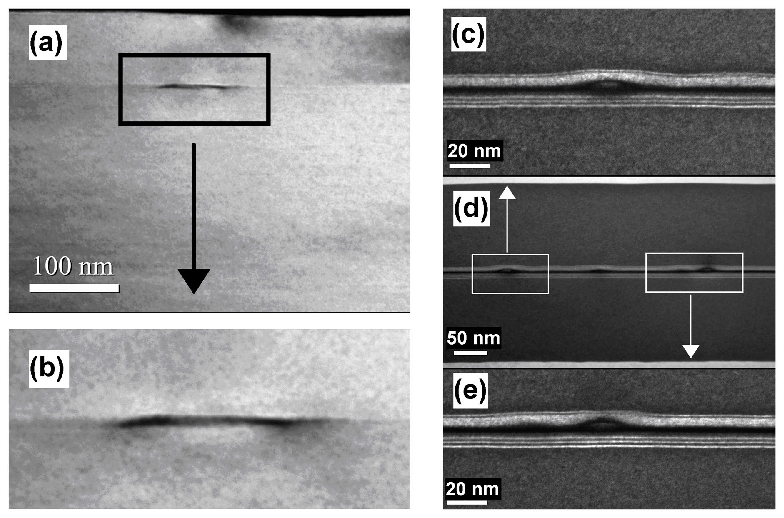}
\caption{\label{fig:TEM} Transmission electron microscope (TEM)
images showing the structure of InP/GaInP (a-b) and InGaAs/GaAs
(c-e) QDs. (b) shows zoomed part of the image in (a), (c) and (e)
are zoomed parts of (d).}
\end{figure}

The InGaAs/GaAs sample \cite{InGaAsSamp1,InGaAsSamp2,InGaAsSamp3}
consists of a single layer of InAs quantum dots (QDs) placed
within a microcavity structure which is used to select and enhance
the photoluminescence from part of the inhomogeneous distribution
of QD energies. The sample has been produced using a single step
MBE process and consists of a GaAs cavity of thickness $\lambda/n$
formed between an asymmetric set of distributed Bragg reflector
pairs, which uses 16 pairs of GaAs/Al$_{0.8}$Ga$_{0.2}$As
distributed Bragg reflector pairs below and 6 pairs above the
cavity. The resulting cavity Q factor is $\sim$250 and the cavity
has a low temperature resonant wavelength at around 920~nm. The
luminescence of the QDs is further enhanced by the presence of a
GaAs/AlAs short-period superlattice surrounding the QD layer. The
superlattice and DBR layers are grown around 620$^{\circ}$~C. The
quantum dots were formed by deposition of ~1.85~monolayers (MLs)
of InAs at a growth temperature of 510$^{\circ}$~C and a growth
rate of 0.1 ML/s. The deposition amount is just above that
required for the nucleation of QDs ($\sim$1.65~ML) but is well
below the amount required to produce a mature QD distribution. As
a result, we obtain a low density of infant QDs at the
post-nucleation stage, which are small and have a low
concentration of indium.

Both QD samples were examined using transmission electron
microscopy (TEM). Fig. \ref{fig:TEM} shows images for InP/GaInP
(a-b) and InGaAs/GaAs (c-e) samples taken under dark field 002
condition, which gives compositionally-sensitive diffraction
contrast \cite{JMI:BEANLAND}. We find that InP dots are disk
shaped with lateral size of $\sim$75~nm and a height of
$\sim$4.5~nm. Images taken on the InGaAs/GaAs sample clearly show
the AlGaAs layers of the Bragg mirror and the superlattice as well
as pyramidal shaped QDs approximately 30~nm wide at their base and
5~nm high in the center.

\section{\label{SI:Experiment}Details of experimental techniques}

\begin{figure}
\includegraphics{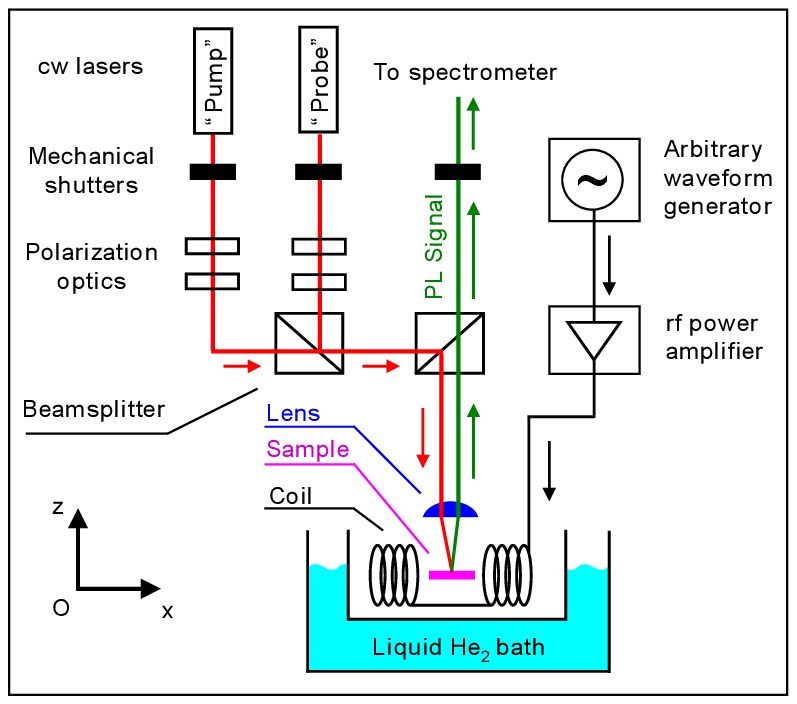}
\caption{\label{fig:Setup} Experimental setup. The sample is
placed inside an insert filled with low-pressure helium gas with
optical access from the top. The insert is immersed in a liquid
helium bath cryostat. Photoluminescence (PL) is excited by diode
lasers. All experiments are performed in the Faraday geometry with
excitation and collection normal to the sample surface. Focusing
of the laser and collection of PL from a QD are performed with an
aspheric lens with focal length $f\approx2$~mm and numerical
aperture NA$\approx0.5$. The collected PL signal is coupled via an
optical fiber to the entrance slit of a double spectrometer with a
charged coupled device (CCD) camera. Using line fitting or
calculating center of mass it is possible to determine the shifts
in QD PL energy as small as 1~$\mu$eV. We use two identical lasers
combined via a beamsplitter which allow independent control of
power and polarization of ''pump'' and ''probe'' pulses.
Mechanical shutters are used to control the light of both lasers
and PL signal according to the pump-probe timing diagram [Fig.
\ref{fig:SpecScheme}(c) of the main text]. Time accuracy of the
shutters is $\sim$1~ms. Magnetic field up to 8~T can be applied
perpendicular to the sample surface (along the $Oz$ direction).
Oscillating radio-frequency (rf) magnetic field along the $Ox$
direction is produced by two coils on each side of the sample. The
rf signal is produced by digital arbitrary waveform generators
with analog modulators, and after the power amplifier, coupled to
the coil via a coaxial cable.}
\end{figure}

The experiments are performed with the sample placed in an
exchange-gas cryostat at $T=4.2$~K, and using an external magnetic
field $B_z$ normal to the sample surface. In order to detect
nuclear polarization on the dot we use high resolution
micro-photoluminescence (PL) spectroscopy of single QDs (see
experimental setup scheme in Fig. \ref{fig:Setup}). The QD PL is
excited by a laser resonant with the wetting layer states
($E_{exc}$=1.88~eV for InP dots and $E_{exc}$=1.46~eV for InGaAs
dots) and analyzed with a 1~meter double spectrometer coupled to a
CCD. Manipulation and probing of nuclear spin polarization relies
on the hyperfine interaction of electrons and nuclear spins
\cite{Tartakovskii}, and requires polarization-resolved excitation
and detection of light as described in the main text.

\begin{figure}
\includegraphics{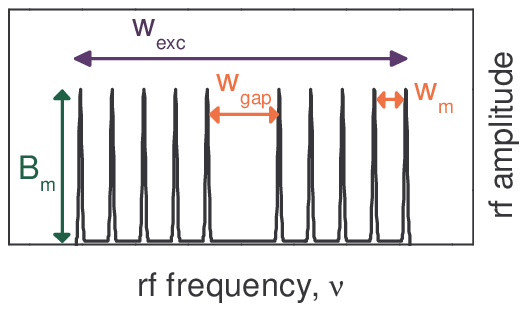}
\caption{\label{fig:RFSpecSup} Schematics of the spectrum of rf
excitation. The spectrum has a total width $w_{exc}$ up to 20 MHz
and consists of many modes with equal amplitudes $B_m$ and spacing
$w_m$, which is varied in different measurements in the range
$0.1\div4$ kHz ($w_m=0.4$~kHz is used for ''inverse''
spectroscopy). For ''inverse'' NMR spectroscopy the spectrum can
also have a gap in the center with a width $w_{gap}$ varied in the
range 8-800 kHz.}
\end{figure}

The waveform for the radio-frequency excitation is produced by a
digital arbitrary waveform generator. ''Inverse'' spectroscopy
with a continuous rf excitation spectrum shown in Fig.
\ref{fig:InvNMRScheme}(d) requires an aperiodic signal, which can
not be produced by a digital device. We thus approximate the
spectrum shown in Fig. \ref{fig:InvNMRScheme}(d) with a spectrum
consisting of a large number of discrete modes with equal spacing
$w_m$ (typically $w_m=0.4$~kHz) as shown in Fig.
\ref{fig:RFSpecSup}. Since the rf signal has a finite power, it
has to be limited in the spectral domain to a width $w_{exc}$
(typically $w_{exc}=20$~MHz). The signal waveform is synthesized
to have no spectral components within a gap of width $w_{gap}$.
However due to imperfections of the rf circuit (harmonics and
spurious noise) the amplitude of the rf field within the gap is
not strictly zero. In our experiments the spectral density of the
rf power inside the gap is $\sim$1000 times smaller than the
spectral power density of the modes. The typical mean square
amplitude of the in-plane rf oscillating magnetic field used in
the ''inverse'' spectroscopy experiments is $\sqrt{\langle
B^2_{rf}\rangle}\approx0.15$~mT, while the phases of the modes are
chosen to minimize the crest factor so that the peak value is
$B_{rf}\lesssim0.3$~mT. In our experimental setup continuous rf
excitation of such amplitude results in sample heating of less
than 1~K.

The duration of the rf pulse for ''inverse'' spectroscopy
$T_{rf}=5.5$~s is chosen to be long enough to produce the
steady-state population probability distribution of the nuclear
spin states shown in Fig. \ref{fig:InvNMRScheme}(f). On the other
hand for spectroscopy on phosphorus we use $T_{rf}$=50~ms which is
shorter than the time required to completely erase nuclear
polarization, and thus gives an unsaturated spectrum allowing
broadening to be avoided. Both InP and InGaAs dots used in this
work exhibit long nuclear spin decay times \cite{InPDyn}
$T_1>$100~s so that natural decay of nuclear polarization during
rf excitation and probe pulse is negligible.

The ''inverse'' spectra (see Fig. \ref{fig:NMRSpec}) measured with
$\sigma^+$ optical pumping were calculated as a difference
$E_z^{gap}-E_z^{no\; gap}$ of the exciton Zeeman splitting
$E_z^{gap}$ measured using rf excitation with a gap in the
spectrum at frequency $\nu$ and the splitting $E_z^{no\; gap}$
measured without the gap. Such correction allows NMR signal to be
expressed as an absolute value of the hyperfine spectral shift.
For $\sigma^-$ pumping the difference was taken with the opposite
sign $(E_z^{no\; gap}-E_z^{gap})$ to simplify comparison with
spectra measured under $\sigma^+$ pumping. For ''saturation''
spectroscopy $\sigma^+$ pumping was used and the signal was
calculated as $E_z^{no\; rf}-E_z^{rf}$, where
$E_z^{rf}$($E_z^{no\; rf}$) is the splitting measured with
(without) rf excitation centered at frequency $\nu$.

\section{\label{SI:NucSpecStrain}Nuclear spin spectrum in the presence of strain}

Below we briefly summarize the effect of external fields on the
nuclear spin spectrum. In magnetic field $B_z$ along the $Oz$ axis
and in the presence of elastic strain, the Hamiltonian for a
nuclear spin $I$ can be written as \cite{AbrahamBook}:
\begin{eqnarray}
H_{nuc}=-h\nu_L I_z+H_Q,\label{eq:Hnuc}
\end{eqnarray}
where $\nu_L=\gamma B_z/(2\pi)$ is Larmor frequency, $h$ - Planck
constant, $\gamma$ is nuclear gyromagnetic ratio and $H_Q$
describes interaction of the nuclear quadrupole moment with the
electric field gradient (EFG), described by a second rank
traceless tensor of the electrostatic potential second derivatives
$V_{ij}$. In the frame $Ox'y'z'$ with the axes along the principal
axes of $V_{ij}$:
\begin{eqnarray}
H_{Q}=h\nu_Q
\left[3I_{z'}^2-I^2-\eta(I_{x'}^2-I_{y'}^2)\right]/6,\label{eq:HQ}
\end{eqnarray}
where $\nu_Q=\frac{3 e Q V_{z'z'}}{2I(2I-1)h}$ and
$\eta=\frac{V_{x'x'}-V_{y'y'}}{V_{z'z'}}$ describe strength and
deviation of the EFG from the axial symmetry, respectively
\cite{AbrahamBook}.

\begin{figure}
\includegraphics{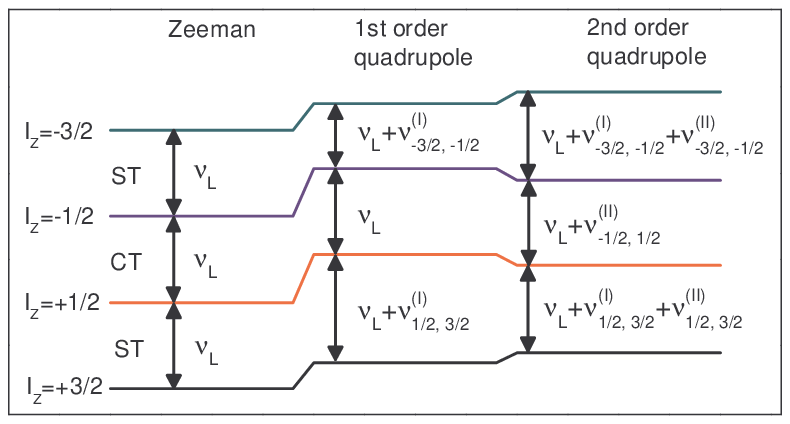}
\caption{\label{fig:NucLevels} Diagram of nuclear spin energy
levels for the case of $I$=3/2 spin. Dipole transitions are shown
with arrows. Zeeman interaction leads to equidistant shift of spin
levels resulting in a degenerate transition at a Larmor frequency
$\nu_L$. Quadrupole interaction, which can be treated as a
perturbation, leads to further shifts. The first order correction
makes transition frequency $\nu_L+\nu^{(I)}_{I_z',I_z'+1}$
dependent on the spins of the levels $I_z'$ and $I_z'+1$ thus
removing the degeneracy. However the correction
$\nu^{(II)}_{-1/2,1/2}$ to the frequency of the central transition
(transition between -1/2 and 1/2 states) appears only in the
second order.}
\end{figure}

In the case of high magnetic field ($\nu_L\gg \nu_Q$) studied here
the quadrupole interaction can be treated as a perturbation. The
effect of external fields on nuclear spin level energies for the
case of spin $I$=3/2 is shown in Fig. \ref{fig:NucLevels}. Without
quadrupole interaction frequencies of all transitions equal to
$\nu_L$ determined by Zeeman energy. In the first order of
$\nu_Q$, the frequency of the $I_z'\leftrightarrow I_z'+1$
transition, $\nu_L+\nu^{(I)}_{I_z',I_z'+1}$, becomes dependent on
$I_z'$. For the uniaxial EFG with the main axis along the external
magnetic field this shift reads as
$\nu^{(I)}_{I_z',I_z'+1}=\nu_Q(I_z'+1/2)$, i.e. $2I$ NMR lines are
observed equally spaced by $\nu_Q$. In the case of non-axial
symmetry of the EFG ($\eta\ne0$) or non-zero angle $\theta$
between EFG principle axis $Oz'$ and magnetic field direction
$Oz$, further changes in the transition frequencies are observed
and the first order shift reads as:
\begin{eqnarray}
\nu^{(I)}_{I_z',I_z'+1}=\nu_Q (I_z'+1/2)
(1+3\cos2\theta-2\eta\cos2\phi\sin^2\theta)/4,\label{eq:nuPT1}
\end{eqnarray}
where $\phi$ is the angle describing orientation of EFG axes with
respect to $Ozz'$ plane. For example, it follows from Eq.
\ref{eq:nuPT1} that for $\eta=0$ the first-order quadrupole shift
can be canceled if strain axis deviates from the magnetic field by
an angle $\theta\approx54.7^{\circ}$.

However, the shift of the central transition
$-1/2\leftrightarrow1/2$ appears only in the second order of
$\nu_Q$ and can be written as:
\begin{eqnarray}
\nu^{(II)}_{-1/2,1/2}=\frac{2}{9}\frac{\nu_Q^2}{\nu_L}
(I(I+1)-3/4) G(\theta,\eta,\phi),\label{eq:nuPT2}
\end{eqnarray}
where $G(\theta=0,\eta=0,\phi)=0$ and $-1<G(\theta,\eta,\phi)<1/2$
for all possible values of $\theta,\eta,\phi$ [complete
expressions for $G(\theta,\eta,\phi)$ can be found in Ref.
\cite{QEncycl}]. Thus for $\nu_L\gg \nu_Q$ the shift of the CT is
much smaller than for the STs, resulting in a narrow central peak
in the NMR spectra.

\section{\label{SI:InvNMR}Calculation of the NMR signal for ''inverse''
spectroscopy}

Let us first consider ''saturation'' spectroscopy on quadrupolar
nuclei subject to EFG when the spectrum of rf excitation has a
single frequency component only. If the rf frequency $\nu$ is
equal to the frequency of the transition between $I_z'$ and
$I_z'+1$ states [$-3/2\leftrightarrow -1/2$ in Figs.
\ref{fig:InvNMRScheme}(a-c)], the populations of these spin levels
are equalized [the dotted lines in Fig.
\ref{fig:InvNMRScheme}(c)]. The populations of all other spin
levels are not affected, and the total change of nuclear spin
polarization $P_N$ due to rf excitation is determined only by the
contributions of $I_z'$ and $I_z'+1$ states. Taking the difference
between the initial polarization $[(I_z'+1)p_{I_z'+1}+I_z'
p_{I_z'}]/I$ and the polarization after saturation of the
transition $[(I_z'+1)+I_z']\times(p_{I_z'+1}+p_{I_z'})/(2I)$, we
find the amplitude of the detected NMR signal as:
\begin{eqnarray}
\Delta P_N^{sat}=(p_{I'_z+1}-p_{I'_z})/(2I). \label{eq:SatNMR}
\end{eqnarray}
By contrast in the ''inverse'' spectroscopy method [Figs.
\ref{fig:InvNMRScheme}(d)-(f)] all spin states contribute to the
NMR signal. For simplicity we consider the case of large EFG and
narrow gap in the rf spectrum, i. e. when the frequency of only
one nuclear transition coincides with the gap. Calculations for a
general case are bulky but straightforward. First we note that if
the gap is not in resonance with any transitions then all nuclear
polarization will be erased yielding the final polarization
$P_N=0$. If the gap is in resonance with the $I_z'\leftrightarrow
I_z'+1$ transition [$-3/2\leftrightarrow -1/2$ in Figs.
\ref{fig:InvNMRScheme} (d-f)], i. e.  this transition is not
excited, then the equalization of populations takes place
separately for two groups of spin levels: for states with $I_z\le
I_z'$ and for states $I_z\ge I_z'+1$ [dotted lines in Fig.
\ref{fig:InvNMRScheme}(f)]. This is because transfer of population
induced by the rf field is allowed only between spin levels with
$I_z$ differing by $\pm1$. For each group, the population
probabilities of the spin levels after the rf pulse will be the
average of the their initial populations $p_{I_z}$. Thus non-zero
polarization $\Delta P_N^{inv}$ will be retained:
\begin{eqnarray} \Delta
P_N^{inv}=\frac{I_z'-I}{2 I}\sum_{k=-I}^{I_z'} p_{k}+
\frac{I+I_z'+1}{2 I}\sum_{k=I_z'+1}^{I} p_{k}.\label{eq:InvNMR}
\end{eqnarray}

The initial polarization of the nuclear spins created via the
hyperfine interaction with the optically pumped spin-polarized
electrons can be expressed in terms of the electron spin
temperature $T_e$, assuming that the nuclear spin populations
$p_{I_z}$ follow a Boltzman distribution. In large magnetic fields
when $\nu_L\gg\nu_Q$, the energies of the nuclear spin levels are
approximately proportional to $I_z$ and the initial populations
can be described as \cite{DyakonovPerel1973}:
\begin{eqnarray}p_{I_z}=Z^{-1}\exp(I_z \beta),\nonumber\\
\beta=\left(\frac{E_{eZ}}{k_B T}-\frac{E_{eZ}}{k_B
T_e}\right),\label{eq:BoltzmanDistr}
\end{eqnarray}
where $Z$ is the normalization factor, $k_B$ - Boltzman constant,
$E_{eZ}$ - the electron Zeeman splitting. $\beta$ describes the
dynamic nuclear polarization occurring as a result of deviation of
the electron spin temperature $T_e$ from the bath temperature $T$
due to optical orientation of the electrons. In the studied case
$|T_e|\ll T$ and the term $E_{eZ}/(k_B T)$ can be neglected.

Using the Boltzman population distribution of Eq.
\ref{eq:BoltzmanDistr} we can calculate the sums in Eqs.
\ref{eq:SatNMR} and \ref{eq:InvNMR} and obtain the following
expressions for the signal amplitudes for both ''saturation'' and
''inverse'' NMR:
\begin{eqnarray}\Delta
P_N^{sat,B}=
\frac{[e^{(I_z'+1)\beta}-e^{I_z'\beta}]}{2I}\frac{\sinh(\beta/2)}{\sinh[(2I+1)\beta/2]},\nonumber\\
\Delta P_N^{inv,B}=
\frac{(I-I_z')+(I+I_z'+1)e^{(2I+1)\beta}-(2I+1)e^{(I+I_z'+1)\beta}}{2I[e^{(2I+1)\beta}-1]}.
\label{eq:InvNMRBoltz}
\end{eqnarray}
It is useful to consider several practical cases. For example, the
amplitude of the $-1/2\leftrightarrow 1/2$ central transition (CT)
signal is obtained from Eqs. \ref{eq:InvNMRBoltz} by substituting
$I_z'=-1/2$. For a given $I$, the CT NMR signal $\Delta P_N$
becomes a function of $\beta$ and it can be easily derived that
for any $\beta$, ''inverse'' spectroscopy yields a signal
enhancement of at least $\Delta P_N^{inv,B}/\Delta
P_N^{sat,B}>(I+1/2)^3$. For example, for $I=9/2$ spin (indium) it
is $\Delta P_N^{inv,B}/\Delta P_N^{sat,B}>125$. Since the CT
exhibits a narrow line in the NMR spectrum it is much easier to
detect even in the presence of large quadrupole effects. Thus the
use of ''inverse'' NMR greatly enhances the sensitivity for
detection of even small amounts of quadrupole isotopes, in
particular with large $I$. For the smallest half-integer
quadrupole nuclear spin $I=3/2$, ''inverse'' NMR for the CT still
leads to a signal enhancement of $\Delta P_N^{inv,B}/\Delta
P_N^{sat,B}>8$. For the satellite transitions (STs) the
sensitivity of ''inverse'' NMR reduces with increasing $I_z'$.
However, even for the most split-off transitions
$-I\leftrightarrow -I+1$ and $I-1\leftrightarrow I$, the
enhancement is $\Delta P_N^{inv,B}/\Delta P_N^{sat,B}>9$ for
$I=9/2$ and $>3$ for $I=3/2$.

It is also useful to compare NMR sensitivity in strained and
unstrained structures. In the absence of quadrupole effects all
nuclear transitions have the same frequencies, and thus for both
''saturation'' and ''inverse'' techniques the NMR signal will be
given by the initial polarization degree $P_N$. Using Eqs.
\ref{eq:BoltzmanDistr} and \ref{eq:InvNMRBoltz} we can calculate
the ratio of the CT NMR signal $\Delta P_N$ in the presence of EFG
and the total signal $P_N$ for zero EFG. For ''inverse'' NMR we
find that for all $\beta$ this ratio is $\Delta
P_N^{inv,B}/P_N>0.55$ for $I=9/2$ ($\Delta P_N^{inv,B}/P_N>0.66$
for $I=3/2$), implying that sensitivity in the strained structures
is only two times smaller compared to the structures with zero
quadrupole effect. By contrast, for saturation spectroscopy,
non-zero EFG results in suppression of NMR signal particularly
strong for nuclei with large spin: we find that at least $\Delta
P_N^{sat,B}/P_N<0.0061$ for $I=9/2$ and $\Delta
P_N^{sat,B}/P_N<0.1$ for $I=3/2$.

\section{\label{SI:ChemComp}Estimation of the chemical composition of the dots}

This section presents an experimental method and numerical
analysis enabling estimation of Ga and In intermixing within the
volume of the electron wavefunction in a QD. We use a long rf
pulse leading to selective (and complete) depolarization of the
$i$-th isotope, thus enabling a selective measurement of the
corresponding hyperfine shift (the change in the exciton spectral
splitting) $\Delta E_{Z,i}$. The $\Delta E_{Z,i}$ can be expressed
via the nuclear polarization degree $P_{N,i}$ of the $i$-th
isotope as:
\begin{eqnarray}
\Delta E_{Z,i}=\rho_i A_i I P_{N,i}, \label{eq:hfShift}
\end{eqnarray}
where $\rho_i$ is the relative concentration of that isotope and
$A_i$ is its hyperfine constant \cite{AbrahamBook}. For InP, where
the contribution of $^{115}$In dominates, the hyperfine constant
has been measured experimentally \cite{Gotschy}:
$A_{^{115}In}\approx 47$~$\mu$eV. Using this value, and neglecting
variation of electron density between gallium and indium sites, we
can estimate the hyperfine constant for $^{69}$Ga as
$A_{^{69}Ga}=A_{(^{115}In)}\frac{\mu_{(^{69}Ga)}}{\mu_{(^{115}In)}}\frac{9/2}{3/2}\approx
51$~$\mu$eV, and similarly for $^{71}$Ga as $A_{^{71}Ga}\approx
65$~$\mu$eV. We also take into account that the two gallium
isotopes have natural abundances $\sigma_{^{69}Ga}\approx 0.6$ and
$\sigma_{^{71}Ga}\approx 0.4$. By introducing the total gallium
concentration $\rho_{Ga}$ we can write
$\rho_{^{69}Ga}=\sigma_{(^{69}Ga)}\rho_{Ga}$,
$\rho_{^{71}Ga}=\sigma_{(^{71}Ga)}\rho_{Ga}$. Since $^{115}$In and
$^{69}$Ga have similar NMR frequencies, we can only measure their
combined Overhauser shift $\Delta E_{Z,^{69}Ga}+\Delta
E_{Z,^{115}In}$. The Overhauser shift for $^{71}$Ga, $\Delta
E_{Z,^{71}Ga}$, is measured separately. Finally, the nuclear spin
polarization for each isotope $P_{N,i}$ can be calculated using
the Boltzman distribution (Eq. \ref{eq:BoltzmanDistr}) and thus
expressed in terms of polarization parameter $\beta$ for a given
spin $I$. Using Eq. \ref{eq:hfShift} we can write the following
system of equations for $\rho_{Ga}$, $\rho_{In}$ and $\beta$:
\begin{eqnarray}
\Delta E_{Z,^{69}Ga}+\Delta E_{Z,^{115}In}=\frac{9}{2}\rho_{In}
A_{(^{115}In)} P_{N,(^{115}In)}+\frac{3}{2}\sigma_{(^{69}Ga)}\rho_{Ga}A_{(^{69}Ga)}P_{N,(^{69}Ga)},\nonumber\\
\Delta E_{Z,^{71}Ga}=\frac{3}{2}\sigma_{(^{71}Ga)}\rho_{Ga}A_{(^{71}Ga)}P_{N,(^{71}Ga)},\nonumber\\
\rho_{Ga}+\rho_{In}=1. \label{eq:hfEq}
\end{eqnarray}
For InP/GaInP QDs we measured the following values of the
hyperfine shifts: $\Delta E_{Z,^{115}In}+\Delta
E_{Z,^{69}Ga}\approx120$~$\mu$eV and $\Delta
E_{Z,^{71}Ga}\approx8$~$\mu$eV. Solving Eq. \ref{eq:hfEq} we find
$\rho_{Ga}\approx35\%$, $\rho_{In}\approx65\%$ implying
significant penetration of the electron wavefunction into the
GaInP barrier and/or diffusion of gallium into the dot. We also
find $\beta\approx0.8$ corresponding to the average electron spin
of $|\langle s_z \rangle|\approx0.2$. For InGaAs quantum dots we
assume the same values of hyperfine constants $A_{i}$. Using the
measured shifts $\Delta E_{Z,^{115}In}+\Delta
E_{Z,^{69}Ga}\approx56$~$\mu$eV and $\Delta
E_{Z,^{71}Ga}\approx18$~$\mu$eV, we find $\beta\approx0.8$ and, as
expected from peaks amplitudes of the NMR spectrum in Fig.
\ref{fig:NMRSpec}(c), a much lower concentration of indium
$\rho_{In}\approx20\%$. For both types of quantum dots we find
very similar high degrees of optically pumped nuclear spin
polarization: $P_{N,(^{115}In)}\approx0.8$ and
$P_{N,(^{71}Ga)}\approx0.6$.

There are two possible scenarios resulting in relatively low
average concentration of indium ($\rho_{In}\approx20\%$) in InGaAs
QDs revealed by the NMR measurements: (i) the nominal InAs QD
contains significant amount of gallium due to diffusion during the
growth process, or (ii) the dot itself consists mainly of indium
but has a small size resulting in weak charge localization and
significant penetration of electron wavefunction into the GaAs
barrier. In both cases QDs will have a larger band-gap compared to
the large-size InAs dots. This is indeed observed in the
experiment: the wavelength of the ground state exciton
recombination in the studied dots is $\lambda\sim915$~nm compared
to $\lambda > 1000$~nm in large indium-abundant dots.

\section{\label{SI:StrainEstim}Estimation of strain in a QD}

Elastic strain is described by a second-rank symmetric tensor
$\epsilon_{ij}$. The resulting electric field gradient (EFG)
$V_{ij}$ can be related to $\epsilon_{ij}$ via a fourth-rank
tensor $S_{ijkl}$ as \cite{SundforsSTensor}
\begin{eqnarray}
V_{ij}=\sum_{k,l}S_{ijkl}\epsilon_{kl},\quad(i,j,k,l=x,y,z).\label{eq:STensor}
\end{eqnarray}
In a crystal with cubic symmetry there are only 3 independent
components denoted as $S_{11}$, $S_{12}$ and $S_{44}$. The
symmetry relation $S_{12}=-S_{11}/2$ is usually used to account
for the zero trace of the $V_{ij}$ tensor, leaving only 2
independent components \cite{SundforsSTensor}. Thus Eq.
\ref{eq:STensor} can be rewritten as
\begin{eqnarray}
V_{ii}=S_{11}(\epsilon_{ii}-(\epsilon_{jj}+\epsilon_{kk})/2),\quad
i\ne j\ne k\nonumber\\
V_{ij}=2S_{44}\epsilon_{ij},\quad i\ne j. \label{eq:STensorCubic}
\end{eqnarray}
It follows from Eq. \ref{eq:STensorCubic} that isotropic
(hydrostatic) strain
$\epsilon_h=\epsilon_{xx}+\epsilon_{yy}+\epsilon_{zz}$ produces no
EFG, while biaxial (including uniaxial) and shear strains give
rise to quadrupole shifts of NMR resonances.

For uniaxial ($\epsilon_{xx}=\epsilon_{yy}$) strain
$\epsilon_b=\epsilon_{zz}-(\epsilon_{xx}+\epsilon_{yy})/2$ along
the direction of the applied magnetic field, the frequency shift
of the $1/2\leftrightarrow3/2$ transition from the CT reads as
$\nu_Q=\frac{3eQS_{11}\epsilon_b}{2hI(2I-1)}$, where $Q$ is the
quadrupole moment, $h$ is the Planck constant. For the studied
isotopes the quadrupole moments are: $Q(^{69}$Ga$)\approx0.17$,
$Q(^{71}$Ga$)\approx0.10$, $Q(^{75}$As$)\approx0.31$ and
$Q(^{115}$In$)\approx0.8$ barn (1 barn = $10^{-28}$ m$^2$). For
$^{71}$Ga in InGaAs dots the maximum $\nu_Q\sim$2.5~MHz can be
estimated from the width of the sidebands in the NMR spectrum in
Fig. \ref{fig:NMRSpec}(e). Using the value
$S_{11}\approx2.7\times10^{22}$~V/m$^2$ measured for gallium in
bulk GaAs\cite{SundforsSTensor} we estimate
$|\epsilon_b|\approx6\%$.

In InP dots we use NMR on $^{115}$In satellite transitions to
estimate strain magnitude. According to Eq. \ref{eq:InvNMRBoltz},
the amplitude of the ''inverse'' NMR signal from the satellite
transition $I_z'\leftrightarrow I_z'+1$ decreases with increasing
spin $|I_z'|$, leading to insufficient signal for  large $|I_z'|$
and unreliable estimation of the maximum quadrupole shift. This is
overcome in an additional experiment, where long broadband rf
pulse (without the gap) centered at the frequency of the indium CT
transition is used to erase polarization of indium. The total
width of the excitation spectrum $w_{exc}$ is varied. We find that
the magnitude of the erased nuclear polarization initially
increases with $w_{exc}$ and saturates at a constant level at
$w_{exc}\approx20$~MHz. This allows the maximum shift of the
$7/2\leftrightarrow 9/2$ transition frequency to be estimated as
$\sim10$~MHz, which is also equal to $4\nu_Q$. Using then
$S_{11}\approx5.9\times10^{22}$~V/m$^2$ for indium in InP we
estimate the maximum strain as $|\epsilon_b|\approx5\%$.

Elastic strain also affects the CT frequency. Using
$\nu_Q\approx$2.5~MHz derived for $^{115}$In we find that the
shift of $^{115}$In resonance according to Eq. \ref{eq:nuPT2} can
be as large as $\nu^{(II)}_{-1/2,1/2}\approx-0.65$~MHz at
$\nu_L\approx49.7$~MHz. For example, in the case of $\eta=0$ and
angle $\theta\approx54.7^{\circ}$ corresponding to zero shift of
ST bands the second-order shift of the CT is
$\nu^{(II)}_{-1/2,1/2}\approx-0.25$~MHz. On the other hand in
experiment [Fig. \ref{fig:NMRSpec}(b)] we observe shifts only on
the order of $\pm50$~kHz compared to the frequency $\nu_{In}$ in
unstrained InP \cite{BulkInPFreq}. This suggests that the
deviation between the strain axis and external field
(characterized by $\theta$) as well as non-axial symmetry of EFG
(characterized by $\eta$) are small. Thus, the most likely reason
for the inhomogeneous distribution of ST shifts is the variation
of $\nu_Q$ within the dot volume caused by variation of the
biaxial strain magnitude $\epsilon_b$. In particular, non-zero
amplitude of ST bands at CT frequency observed in Fig.
\ref{fig:NMRSpec}(a) for $^{115}$In can be explained by complete
biaxial strain relaxation at the center of the dot
\cite{GrundmannStrain,AndreevStrain} resulting in $\nu_Q=0$. For
$^{75}$As nuclei large EFG can result not only from elastic strain
but also from electric fields created by random substitution of
gallium atoms by indium. This may explain the further broadening
of $^{75}$As CT transition.

We note a significant difference in the asymmetry of the NMR
spectra in Figs. \ref{fig:NMRSpec} (e,f): for $^{75}$As the low
(high) frequency ST band is enhanced for $\sigma^{-(+)}$ pumping
while for indium and gallium isotopes this asymmetry is reversed.
The sign of the asymmetry is determined by the relative signs of
Zeeman and quadrupole contributions to the nuclear Hamiltonian Eq.
\ref{eq:Hnuc}. Since all isotopes of InGaAs have positive
gyromagnetic ratios $\gamma$ and positive quadrupole moments $Q$,
the opposite asymmetries of the spectra can be attributed to the
opposite signs of the electric field gradients $V_{z'z'}$ (see Eq.
\ref{eq:HQ}) experienced by nuclei of anions (As) and cations (Ga
or In).

\section{\label{SI:ModeSpacing}Spin coherence in the nuclear spin ensemble subject to
inhomogeneous strain}

\begin{figure}
\includegraphics{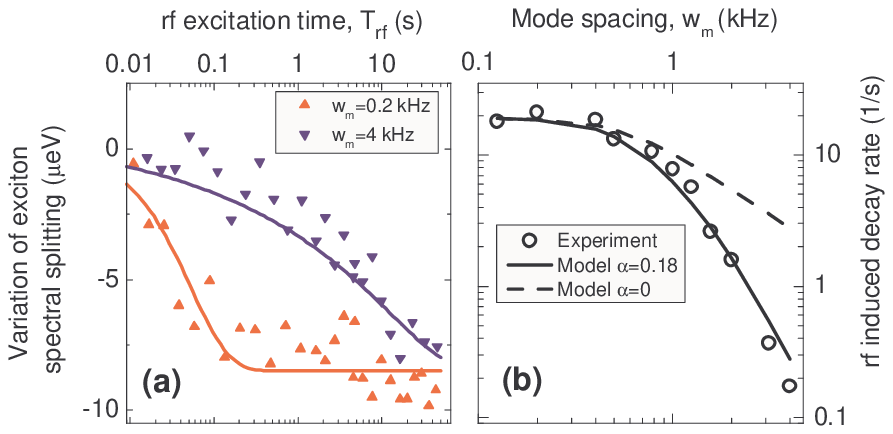}
\caption{\label{fig:ModeSpacingSup}(a) Experimentally measured
nuclear spin polarization decay (symbols) for the rf-excitation
mode spacings $w_m=$0.2~kHz and 4~kHz. Nuclear polarization is
calculated as a difference of exciton spectral splittings $E_Z$
measured after rf-pulse with duration $T_{rf}$ and before this
pulse. Lines show model fitting. (b) Dependence of rf induced
nuclear depolarization rate $R_{rf}$ on the mode spacing $w_m$.
Values extracted using exponential fitting of the experimental
decay curves are shown with symbols. The dashed (solid) line shows
results of calculations using the presented model for a Lorentzian
(non-Lorentzian) lineshape $\alpha=0$ ($\alpha\approx0.18$).}
\end{figure}

This section details the experimental procedure and the model used
for analysis of the dependence of the nuclear polarization
dynamics in InP dots on the spacing between the modes in the
broad-band rf excitation. The dynamics of the rf-induced nuclear
spin polarization decay is measured using a pump-probe technique.
Initially, nuclear spins are polarized optically. Then an rf-pulse
of duration $T_{rf}$ is applied after which the nuclear spin
polarization is probed optically by measuring the exciton Zeeman
splitting $E_z$. The spectrum of the rf excitation (without the
gap) consists of a large number of spectral modes with equal
spacing $w_m$ and with equal amplitudes $B_m$ changed as
$B_m\propto\sqrt{w_m}$ to keep the total power of the rf pulse
constant (See Fig. \ref{fig:ModeSpacing}). The spectral range
44-47.5~MHz of the rf excitation is kept fixed and corresponds to
satellite transitions of In nuclei [Fig. \ref{fig:NMRSpec}(a)].
The experimental dependences of the Overhauser shift on the rf
excitation time $T_{rf}$ are shown with symbols in Fig.
\ref{fig:ModeSpacingSup}(a) for two different mode spacings $w_m$
of the rf-excitation spectrum. As seen, the increase of $w_m$ from
0.2 to 4 kHz results in a significant increase of the time
required to equalize populations of the nuclear spin levels
coupled by the rf field. We use exponential fitting with a time
constant $1/R_{rf}$ to quantify the rate $R_{rf}$ of rf induced
depolarization. The resulting dependence of $R_{rf}$ on $w_m$ is
shown in Fig. \ref{fig:ModeSpacing}(b) and is repeated with
symbols in Fig. \ref{fig:ModeSpacingSup}(b).

We now present in full the model that allows the study of
coherence properties of nuclear spins using nuclear spin dynamic
measurements. We start by noting that, in high external magnetic
fields, nuclear spins are characterized by very long $T_1$ times
\cite{InPDyn} exceeding 100~s, arising from suppressed energy
relaxation leading to low probability of spin flips. Intrinsic
linewidths of NMR transition correspond to much shorter times
$T_2$ determined by energy-conserving decoherence of transverse
components of nuclear spins. This decoherence is induced by random
dipole-dipole fields of other nuclei, and is characterized by the
autocorrelation function which we choose in the form:
\begin{eqnarray}
g(t)=\frac{1}{1-\alpha}\exp[-t/T_2]-\frac{\alpha}{1-\alpha}\exp[-t/(\alpha
T_2)],\label{eq:gCorrelator}
\end{eqnarray}
where $\alpha\ll1$ is a dimensionless parameter. For $\alpha=0$
this corresponds to an exponential correlation function with the
relaxation time $T_2$ derived in the general theory of relaxation
caused by fluctuations \cite{LandauLifshitzStatPhys}. The addition
of the second term in Eq. \ref{eq:gCorrelator} allows the
condition $g'(0)=0$ required by time reversal symmetry to be
satisfied. The lineshape of the nuclear spin transition is found
as the Fourier transform of $g(t)$:
\begin{eqnarray}
\tilde{g}(\nu)=\frac{T_2(1+\alpha)}{1+(1+\alpha)^2 (2\pi \nu)^2
T_2^2+\alpha^2 (2\pi \nu)^4 T_2^4}.\label{eq:LineShape2}
\end{eqnarray}
As expected for $\alpha=0$ this expression corresponds to a
Lorentzian lineshape.

For a nuclear spin transition between $I_z'$ and $I_z'+1$ states
at a frequency $\nu_{nuc}$, a small amplitude (non-saturating) rf
field will result in depolarization, which can be described by the
differential equation for the population probabilities
$d(p_{I_z'+1}-p_{I_z'})/dt=-W\times (p_{I_z'+1}-p_{I_z'})$. For
broad-band rf excitation consisting of discrete modes with
frequency spacing $w_m$, each inducing an equal magnetic field
$B_m$, depolarization rate $W$ is given by:
\begin{eqnarray}
W(\nu_{nuc})=B_m^2\sum_{k=0}^N \tilde{g}(\nu_{nuc}-\nu_0-k\times
w_m),\label{eq:RFDepRate}
\end{eqnarray}
where the summation goes over all modes of the rf field with
frequencies $\nu_k=\nu_0+k\times w_m$ ($\nu_0$ is the frequency of
the first spectral mode). Since the total width of the rf band is
much larger than $w_m$ or the nuclear spin transition width
$w_{nuc}$, the summation in Eq. \ref{eq:RFDepRate} can be extended
to $\pm\infty$.

For each individual pair of nuclear spin states, the rf field will
induce an exponential decay of polarization. However, the QD
contains a large number of nuclear spins with randomly distributed
transition frequencies $\nu_{nuc}$. Thus, in order to calculate
the dynamics of the total nuclear polarization, we need to average
over all possible values of $\nu_{nuc}$. Since the spectrum of the
rf band is a periodic function (i.e. the modes are equally
spaced), such averaging can be done over one period. We also
assume a uniform distribution of the nuclear transition
frequencies $\nu_{nuc}$. Thus, the following expression is
obtained for the time dependence of the difference between the
current and initial nuclear spin polarization, describing the
dynamics of rf-induced depolarization:
\begin{eqnarray}
P_N (t)- P_N (0)=- P_N (0) + P_N (0) \int_0^{w_m}\exp\left(-t B
\sum_{k=-\infty}^{\infty} \tilde{g}(\nu_{nuc}-k\times
w_m)\right)d\nu_{nuc},\label{eq:IRFDec}
\end{eqnarray}
where $B^2\propto B_m^2/w_m$ is proportional to the spectral
density of the total rf field power, which is kept constant in our
experiment when $w_m$ is varied.

Using Eq. \ref{eq:IRFDec} we can fit nuclear spin decay curves
measured for different values of $w_m$. In Fig.
\ref{fig:ModeSpacingSup}(a) this is demonstrated for $w_m$=0.2 and
4 kHz. From the fitting of the whole set of decay curves measured
with rf excitation in the spectral range 44-47.5~MHz for different
$w_m$ we find good agreement with experiment for a nuclear spin
dephasing time $T_2\approx2.5$~ms and lineshape parameter
$\alpha\approx0.18$. The same magnitudes of these parameters are
obtained from fitting the data of a similar experiment but for an
rf excitation band in resonance with a different section of the
satellite transitions at 47.0-49.0~MHz.

We can now use the results of the fitting to analyze coherent
dynamics of indium nuclear spins. We start by noting that for
small $w_m$ all nuclear spin transitions are excited with nearly
the same efficiency (similar to a white-noise excitation)
resulting in an overall decay close to exponential. For larger
spacing, the model given by Eq. \ref{eq:IRFDec} significantly
deviates from the exponential decay: at the initial stage those
transitions which are in resonance with the rf modes quickly
become depolarized, whereas the decay is significantly slower for
off-resonance transitions. For the analysis of the lineshape
parameter $\alpha$ we characterize the rate of the decay given by
Eq. \ref{eq:IRFDec} using the time $t_{1/2}$ where nuclear
polarization decreases to 1/2 of the initial value. The decay rate
can be estimated as $R_{rf}=\log(2)/t_{1/2}$ (in the limit of
small $w_m$ it coincides with exponential decay rate). The
dependence $R_{rf}(w_m)$ calculated in this way from Eq.
\ref{eq:IRFDec} with parameters obtained from the fitting
($T_2\approx2.5$~ms and $\alpha\approx$0.18) is shown in Fig.
\ref{fig:ModeSpacingSup}(b) with a solid line. Calculation with
the same $T_2$ but for $\alpha$=0 is shown with a dashed line.

It can be seen that in the limit of small $w_m$ decay rate does
not depend on $\alpha$ for both $\alpha$=0 and $\alpha$=0.18.
However, at larger mode spacing modeling using pure Lorentzian
lineshape fails to describe the experiment (i.e. it is impossible
to fit simultaneously all decay curves corresponding to different
$w_m$ using Eq. \ref{eq:IRFDec} with $\alpha=0$). This deviation
is described by the fourth order term $\propto \nu^4$ in the
denominator of Eq. \ref{eq:LineShape2}. This term is responsible
for much smaller depolarization rate at large detunings
$\nu\gg1/(\pi T_2)$ in case of non-Lorentzian lineshape. At large
mode spacing $w_m$ most nuclear spin transitions are excited only
via their ''wings'' (i. e. at large $\nu$), and thus the effect of
non-Lorentzian shape becomes more pronounced as seen in Fig.
\ref{fig:ModeSpacingSup}(b). Strong deviation of the nuclear
transition lineshape from Lorentzian function reveals
non-exponential character of nuclear spin decoherence (described
by Eq. \ref{eq:gCorrelator} in our model) and demonstrates the
potential of the presented technique for deeper understanding of
the spin coherence of quadrupole nuclei.

\quad\\

\section*{\label{SI:Acknowledgement}Acknowledgements}

The Gatan Orius digital TEM camera used for TEM imaging of the
samples was funded by Birmingham Science City: Creating and
Characterising Next Generation Advanced Materials, with support
from Advantage West Midlands and part funded by the European
Regional Development Fund A.M.S. would like to thank the Science
City Research Alliance and the HEFCE Strategic Development Fund
for funding Support.


\begin{thebibliography}{10}
\bibitem{BulkInPFreq}
The $^{115}$In resonant frequency corresponding to unstrained InP
is calculated
  as $\nu_{In}$=$\nu_P\times\nu_{In}^{bulk}/\nu_P^{bulk}\approx49633$~kHz using
  $^{31}$P frequency $\nu_P\approx$91605~kHz measured in a QD and frequencies
  $\nu_{i}^{bulk}$ measured for both isotopes in bulk InP \cite{BulkInPNMR}.
\end{thebibliography}
\end{document}